\documentclass[a4paper]{article}

\usepackage[english]{babel}
\usepackage[utf8]{inputenc}
\usepackage{amsmath}
\usepackage{adjustbox}
\usepackage{graphicx}
\usepackage[colorinlistoftodos]{todonotes}
\usepackage[figurename=Fig.]{caption}
\usepackage{listings}
\def\BibTeX{{\rm B\kern-.05em{\sc i\kern-.025em b}\kern-.08em
    T\kern-.1667em\lower.7ex\hbox{E}\kern-.125emX}}
    
\title{Cyber Deception Reactive: TCP Stealth Redirection to On-Demand Honeypots}

\author{Pedro Beltrán López, Pantaleone Nespoli, Manuel Gil Pérez
\thanks{The authors are with the Department of Information and Communications Engineering, University of Murcia, 30100 Murcia, Spain (e-mail:\{pedro.beltranl, pantaleone.nespoli, mgilperez\}@um.es) (Corresponding author: pedro.beltranl@um.es)}}

\date{\today}

\begin{document}
\maketitle

\begin{abstract}
Cybersecurity is developing rapidly, and new methods of defence against attackers are appearing, such as Cyber Deception (CYDEC). CYDEC consists of deceiving the enemy who performs actions without realising that he/she is being deceived. 
This article proposes designing, implementing, and evaluating a deception mechanism based on the stealthy redirection of TCP communications to an on-demand honey server with the same characteristics as the victim asset, i.e., it is a clone. Such a mechanism ensures that the defender fools the attacker, thanks to stealth redirection. In this situation, the attacker will focus on attacking the honey server while enabling the recollection of relevant information to generate threat intelligence. The experiments in different scenarios show how the proposed solution can effectively redirect an attacker to a copied asset on demand, thus protecting the real asset. Finally, the results obtained by evaluating the latency times ensure that the redirection is undetectable by humans and very difficult to detect by a machine.
\end{abstract}

\section{Introduction}
In recent years, the growing threat of cyber attacks against critical infrastructures has prompted an intense search for cutting-edge strategies and techniques to protect these essential assets\footnote{https://securityintelligence.com/news/high-impact-attacks-on-critical-infrastructure-climb-140/}. This trend is due to the constant increase in the sophistication and frequency of these attacks\footnote{https://www.crowdstrike.com/global-threat-report/}, as well as the continuous evolution of critical infrastructures themselves. Against this backdrop of constant challenge, a cutting-edge technique has emerged as a fundamental building block in cyber defence, i.e., Cyber Deception (CYDEC)~\cite{CYDEC}. This innovative strategy offers defenders the ability to deceive attackers to gain valuable information about the attack and, at the same time, defend the assets.

Concretely, CYDEC\footnote{https://itresit.es/cyber-deception/} techniques address a wide range of aspects in the cybersecurity ecosystem such as prevention, detection, and reaction to threats. While recent research has mainly focused on the prevention and detection phases, reaction represents a critical innovation point. The ability to react effectively to a cyber threat is essential to ensure the protection of compromised assets~\cite{Nespoli2022}.

In this context, CYDEC presents itself as a powerful tool. It allows deceiving the attacker and also strengthening protection systems through continuous learning~\cite{learning}, as it stands out as the only mechanism that enables learning. It provides an additional layer of defence by adapting to the changing tactics of cybercriminals. This learning capability contributes significantly to maintaining the integrity of critical infrastructures while deceiving potential attackers, strengthening the cybersecurity system's overall effectiveness.

Due to the great benefits of deception identified above and the need for innovative reaction methods, new techniques based on deception are appearing which, together with each other, form an innovative and intelligent defence that will help one to efficiently and strategically mitigate threats~\cite{tecnicas}.

One of the most promising deception techniques is redirection~\cite{GilPerez2017}. Such a technique allows one to shift the focus of the attack from the victim to another asset monitored and controlled by us, i.e., we are tricking the attacker into believing that the asset he/she is attacking is correct. Once the attacker develops his/her attack on this new asset or environment prepared by us, wasting his/her time and resources, we can log and study all the malicious behaviour generated at the network and host level to gain informative knowledge and improve detection systems.

A key complementary point to the redirection-based deception technique is the environment or asset where the attacker is to be contained. Currently, the Honey-X~\cite{honey} mechanism is used to draw the attackers' attention to them. Nonetheless, our goal is complement this mechanism by simulating the machine that was being primarily attacked. One of the main challenges of this proposal is to store a clone of all the assets as there may be a large expenditure of resources. Therefore, our proposal consist of employing on-demand cloning, i.e., the asset where the attacker is contained is a Honey-X asset based on the victim specifications but created in real-time when it is demanded, thus maintaining an efficient expenditure of resources.

A crucial characteristic in combining these strategies is their stealth capability, fundamental to deceiving an attacker. This feature will be a priority in both TCP connection redirection and honeypot creation. With this point under control, the attacker will not suspect any of the actions carried out and will be successfully deceived, thus accomplishing our mission.

Therefore, this article proposes to design, implement, and evaluate a reaction method based on CYDEC that uses a stealthy TCP redirection to a twin honeypot of the victim generated in real-time. Covering this mechanism is the need for a deception strategy that reacts efficiently and generates continuous learning. Particularly, it is stealthy and efficiently uses the defender resources. Furthermore, a key point of our research has been the publication of the developed code so that the research community can replicate the experiments as well as use the code to add further functionality~\cite{code}.

In the following, a compilation of state-of-the-art on copy on demand and traffic redirection is presented in Section~\ref{related}. In Section~\ref{design}, we explain the design of our system, from the general architecture to how redirection and copy-on-demand are carried out. In Section~\ref{experiments}, we detail the experiments that have been carried out to demonstrate the performance of the proposed system, and finally, in Section~\ref{conclusions}, we review the main conclusions of the work and propose potential next steps.

\section{Related work}
\label{related}
First of all, it is critical to define the main current developments in connection redirection and copy on demand and indicate their main advantages and disadvantages.

\subsection{Network traffic redirection}
First, in~\cite{red1}, a TCP connection transfer mechanism based on Software-Defined Networking (SDN) was proposed. It employs a network data controller to filter and redirect traffic to honeypots. It improves security by enabling transparent traffic control and TCP session replication without being detected by attackers.

In~\cite{red2}, an SDN-based hybrid honeypot architecture was introduced. This architecture consists of four key components: an OpenFlow switch and an SDN controller, which efficiently classify and migrate attack traffic using SDN. The architecture proves to be effective in detecting various network attacks, and test results support its effectiveness.

The work presented in~\cite{red3} proposed an SDN/Network Functions Virtualization (NFV) enabled honeypot system that uses SDN technology to handle covert attack connections. This approach is based on the principle of moving target defence, where the honeypot uses containers to deceive attackers and protect the network. SDN/NFV technologies enable a dynamic response to build a container shell that attracts and captures attackers effectively. The experiments support the effectiveness and efficiency of the proposed honeypot system, demonstrating how SDN can be a valuable tool for network security.

In~\cite{red4}, concentrated on addressing the issue of exchanging TCP connections in diverse network environments by leveraging SDN. It proposed a method that uses an SDN router and an SDN controller to synchronise TCP sessions during the transition, significantly improving content delivery efficiency and TCP connection continuity. Test results indicate that this SDN-based approach outperforms traditional connection-switching methods in speed.

Finally, in~\cite{red5}, a solution to improve efficiency in Content Delivery Networks (CDNs) using SDN was presented. This solution enabled the synchronisation of two separate TCP sessions, allowing the delivery of content from surrogate servers without interrupting the client's active session. It uses OpenFlow to implement this transparent redirection solution. It highlighted the advantage of improving the efficiency of content delivery and the continuity of the client's TCP connection.

\subsection{Copy on demand}
In~\cite{cop1}, Cyber Deception Experimentation System (CDES) was presented, a platform that focuses on creating on-demand and real-time honeypots to improve network security. The authors detailed its design, implementation, and configuration in network environments and presented the results of comprehensive tests that evaluated its performance and effectiveness in detecting and blocking cyber attacks.

In~\cite{cop2}, a Copy-On-Risk (COR) honeypot was introduced to protect cloud infrastructure against attacks. Unlike traditional honeypots, COR-Honeypot creates customised honeypots on demand by cloning real victims and isolating them in controlled environments. The authors described the key technologies used in COR-Honeypot and highlighted its ability to fool attackers into believing they interact with real targets. 

Finally, it was proposed in~\cite{cop3} that a practical solution for implementing honeynets using Virtual Machine (VM) introspection and VM cloning techniques was proposed. The main focus was intrusion detection and extraction of associated malicious software binaries. The solution uses forensic tools and live memory introspection to ensure transparency and resistance to subversion attacks. It also addressed the challenge of hardware requirements, using copy-on-write disks and shared memory and an innovative routing technique that eliminates the need for network reconfiguration after cloning infected hosts. Experiments demonstrated this solution's high efficiency and scalability, which can even benefit from a darknet to enhance its effectiveness.

As we have seen when analysing the different solutions proposed in the literature, no complete solution simultaneously addresses real-time copy-on-demand and stealthy traffic redirection. Some of the analysed articles individually solve both of the abovementioned problems, however, there is a general lack of a full-fledged solution that incorporates both mechanisms. On the other hand, in the on-demand copying proposals analysed, the latency is mentioned just in the collection phase. Nevertheless, none of them performs a comprehensive analysis of latencies. The evaluation of latencies is essential in these techniques due to the need to maintain stealth, as mentioned above. Additionally, it is clear that a stealthy redirection mechanism capable of redirecting a connection already initiated in time is missing. Furthermore, no article can re-establish the redirected connection to the legitimate server when verifying that it is not an attacker, i.e., none of the articles propose a robust solution. In addition, we have shared the code publicly, unlike the vast majority of the articles analysed~\cite{code}.

To address the aforementioned challenges, our solution features the unique ability to provide a comprehensive and robust response. Integrating real-time on-demand copying with stealthy traffic redirection solves the individual challenges identified in the literature. Specifically, it gives defence agencies an essential tool to react effectively to cyber threats. The ability to continuously learn and adapt to changing conditions in the business environment further reinforces the relevance of our solution in constantly improving security and responsiveness. In summary, implementing the proposed CYDEC approach may represent a significant advance in cybersecurity, with substantial benefits for the defence and resilience of organisations.

\section{System design}
\label{design}
Throughout this section, the main design principles of the proposed deception-based system are presented, justifying the decisions made and focusing on the principal benefits.

\subsection{Architecture}

\begin{figure}[hb]
    \begin{adjustbox}{center}
        \includegraphics[width=15cm]{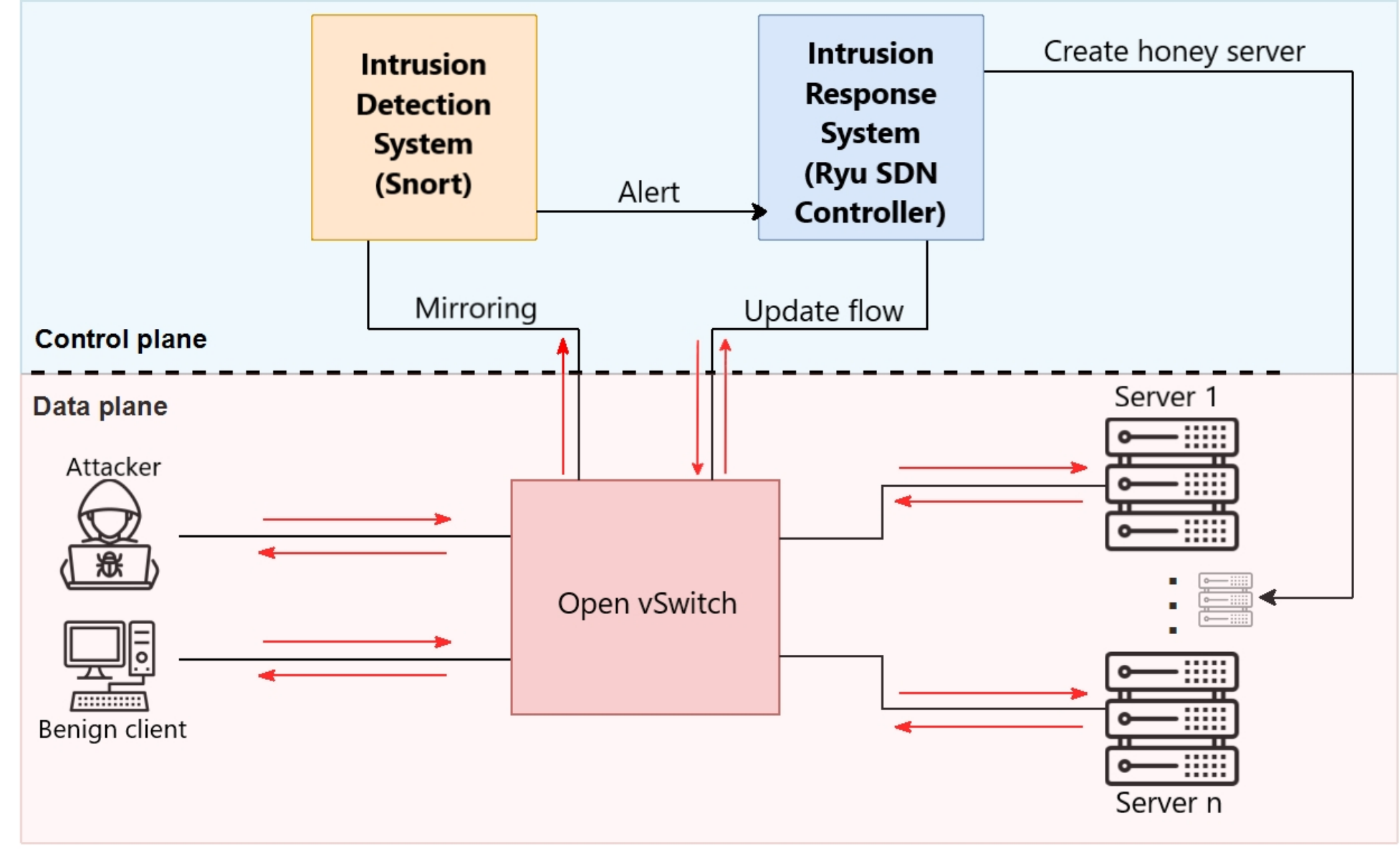}
    \end{adjustbox}
    \caption{Threat detection and response system architecture based on redirection and copy on-demand.}
    \label{cd1}
\end{figure}

The first step in creating a method based on redirection and copying of an asset on demand is designing the architecture to be used. \figurename~\ref{cd1} depicts the high-level architecture of the proposed solution.

Firstly, we have chosen an architecture based on SDN/NFV. This choice relies on the capacity that NFV technology features to host the proposed solution on any type of hardware. On the other hand, SDN offers great versatility in implementing functionalities and defining the actions to be performed for the deception, i.e., message exchange and copy on demand. Concretely, SDNs operate on two different planes, which are described below.

\begin{itemize}
    \item Control plane: It is responsible for creating and maintaining routing tables, allowing the correct functioning of the data plane. In our case, the control plane will be composed of two modules: i) the Intrusion Detection System (IDS), based on Snort software, the IDS scans all network packets for suspicion. According to Snort policies, it has one socket connected to the Switch to receive packets and another connected to the Intrusion Response System (IRS) to send alerts when it detects suspicious activity and ii) the IRS or controller, which will be in charge of carrying out the relevant decisions and actions necessary to route the traffic correctly for CYDEC purposes, it establishes packet exchange rules and updates the routing tables in the Switch to direct traffic. It includes a Ryu controller\footnote{https://ryu-sdn.org/} for deployment and system intelligence and responds to Snort alerts by performing actions such as creating assets on demand or migration.\par

\item Data plane: It contains network packets from clients, servers, and the Open vSwitch (OvS) in our topology. The Switch manages routing according to its routing tables, communicates with the controller for updates, and sends all packets to the IDS for observation. Also belonging to the data plane are the client, the attacker, the servers, and assets of the organisation.
\end{itemize}

\subsection{Redirection implementation}
\label{redirection}
\begin{figure}[!h]
        \begin{adjustbox}{center}
        \includegraphics[width=9cm]{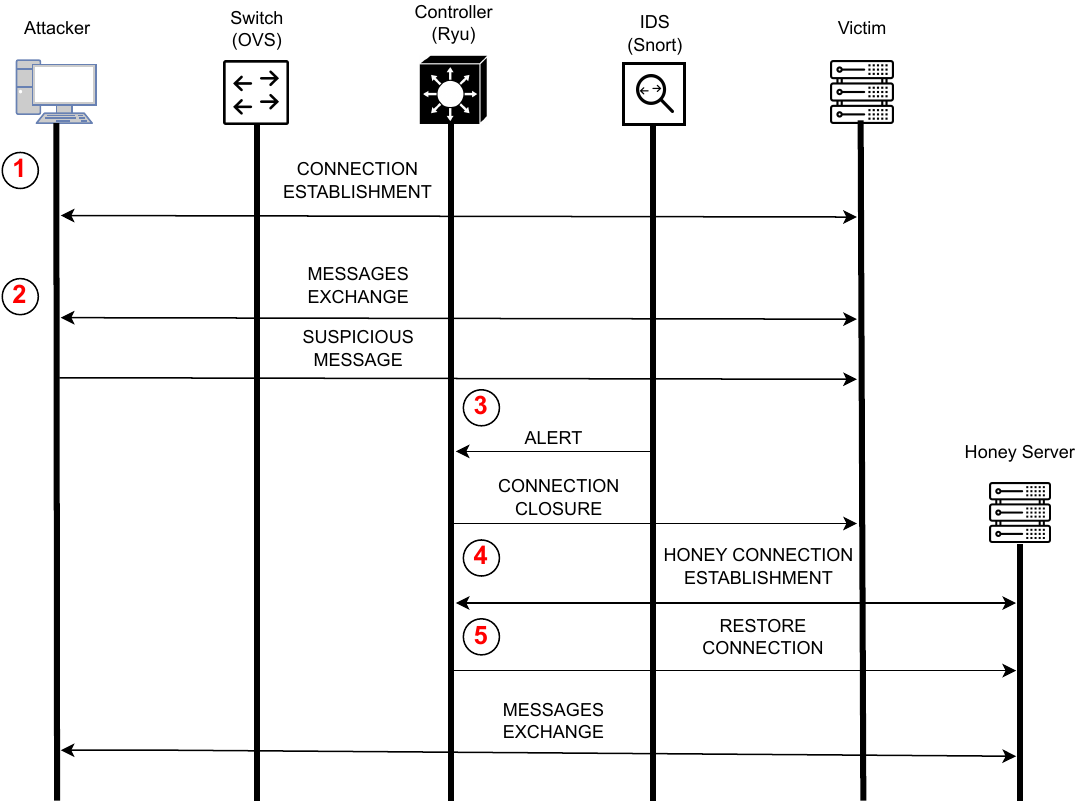}
        \end{adjustbox}
        \caption{TCP packet exchange between the victim and the attacker in the process of stealth redirection.}
        \centering
        \label{cd2}
        \end{figure}

One of the pillars of our system is the creation of a stealthy redirection method capable of protecting the attacked asset instantly and, at the same time, being able to isolate the attacker to study his/her behaviour. Therefore, we will describe how the redirection process is automatically carried out thanks to the IDS alerts, until it exchanges messages with the cloned asset to which the connection has been migrated. These steps can be seen in \figurename~\ref{cd2}:
\begin{enumerate}
    
    \item \textbf{Connection establishment:} The first step in all attacker-server message exchanges is establishing the connection between the attacker and the victim. This connection establishment is carried out with the traditional three-way handshake process.\noindent
   
    \item \textbf{Messages exchange/Suspicious message:} During the exchange of messages between the attacker and the server via the OVS Switch, the attacker at any point in the connection may or may not send a suspicious message, i.e., the IDS has a policy that sends an alert message to the controller.
    
    \item \textbf{Alert/Connection closure:} Once the IDS identifies that the message sent by the attacker is suspicious, it automatically alerts the controller that the message sent by the attacker may be hostile. After, the controller, alerted by the IDS, closes the original server connection to the attacker to protect the server. The attacker does not notice any changes, as his/her connection remains open. At this moment, the copy of the attacked asset (i.e., the honey server) will be created to transfer the connection that has been closed to it.
    
    \item \textbf{Honey connection establishment:} Once the controller has closed the attacker's connection to the victim and the honey server has been instantiated, the establishment of the connection to the honey server begins. This establishment and the subsequent exchange of messages occur while maintaining consistency in the sequence numbers the attacker previously exchanged.
    
    \item \textbf{Restore connection/Messages exchange:} In this step, the exchange of messages between the attacker and the honey server will resume. Finally, once the connection is restored, the attacker exchanges messages with the honey server without realising that the machine is not the victim it was attacking but a clone created in a safe environment.

\end{enumerate}

\subsection{Copy on-demand deployment}
\label{copy}

Establishing a mechanism for copying or cloning an asset and the runtime instantiation of that copy is as important as connection redirection because of the need for stealth in the phase of isolating the attacker so that he/she can carry out his/her objective without suspicion. To carry out this cloning process, several methods can be used. The choice of each method can be determined by two Key Performance Indicators (KPIs) established from our point of view. The first is the time it will take for the cloned asset to be operational in order to continue communicating with the attacker, i.e., from when the order to create a duplicate is issued until it starts communicating with the attacker. The other KPI is the amount of resources consumed by the different mechanisms that will be identified later, i.e., whether these assets will be constantly active, in suspension, or whether they will be created in real-time, consuming resources only when needed efficiently. Therefore, we have identified the following asset cloning mechanisms for subsequent real-time instantiation in our solution:

\begin{itemize}
    \item \textbf{Obtaining information and configuration:} This option can be described as a constant collection of system information and subsequent configuration of the new machine, emulating the configuration of the attacked machine. This process involves triggering periodic service and version recognition mechanisms through our systems. The information collected by one of our assets is used to gain knowledge of the system's status. Maintaining an up-to-date database with the different versions.
 
    \item \textbf{Victim machine image:} The second option is possibly having a repository of updated images of critical machines or those to be protected with the redirection method. It aims to have an image of the attacked machine so that it can be launched at any time with all the services and configurations. This is why, as the configurations of our assets change, the repositories must be updated with the images always to have the latest versions.
    
    \item \textbf{Suspended machine:} Another mechanism involves instantiated at runtime, i.e., to have a copy of the already created asset with minimal resource consumption. When a copy of the asset is needed, it would come out of the suspended state and spend the necessary resources. 
    
    \item \textbf{Copy of the disk where the attacked machine resides:} The last possibility analysed is to copy the disk where the attacked asset resides to create a snapshot copy when the process is activated, i.e., to make a copy of all the machine's information in another space in real-time.
    
\end{itemize}

Once we have identified the different options for cloning the asset, we have to decide which of these options is the best to implement in our solution. However, the choice might differ in different use cases depending on the particular needs. After evaluating several cloning options, the ``Victim machine image'' has been used due to its low latency and minimal resource consumption. Unlike other options with higher latency or the potential to raise suspicion, this choice allows one to keep natural results (latency) on the attacker's end until the creation of the copy is necessary, i.e., the attacker will not be alerted due to the time it takes to create the honey server. This decision is motivated by the critical factor of minimising latency in resuming communication with the attacker, ensuring a discrete approach without alerting the attacker to our actions while minimising the use of resources the defender uses. The choice that is made is set due to the two KPIs specified above, where in our case we want to minimise resource usage and latency. This would be our priority as opposed to any other use case where the priority is different.

\section{Experiments}
\label{experiments}
Once the architecture of the proposed solution has been defined and the system funcionalities and the decisions taken have been detailed, we will carry out two different experiments to demonstrate the operation of the proposed system. The main objective of the following experiments is to demonstrate that it is possible to implement a system capable of successfully deceiving an attacker and to redirect it to a safe environment created in real-time where its behaviour can be studied.

Specifically, Section~\ref{mininet} details the experiment conducted to demonstrate stealth TCP redirection. Meanwhile, Section~\ref{kvm} showcases an experiment illustrating full functionality in a more realistic virtualized environment. In both sections, we provide detailed information about their architectures, configurations, and results.

\subsection{Experiments topology}
To carry out the experiments, the topology shown in \figurename~\ref{a1} has been implemented. In the topology, the different components used are detailed below:

\begin{itemize}
    \item IDS (Snort): The goal of this component is to analyse network traffic to identify any suspicious messages and alert the controller to initiate redirection/copy on demand.
    \item SDN Controller (Ryu): This component is in charge of updating network tables, redirecting, and creating the honey server.
    \item Switch (OvS): The network traffic will be distributed in this component, which directs it from the attacker to the victim and the honey server.
    \item Attacker: The attacker exchanges with the victim machine and sends suspicious messages to alert the IDS.
    \item Victim: This component will host a server that will constantly receive TCP requests from the attacker until the redirection occurs. When the redirection occurs, it will stop receiving packets from the attacker.
    \item Honey server: Like the victim, it will be hosted on the same server, configured similarly, and will start receiving packets once the redirection is complete.
\end{itemize}

\subsection{Redirection}
\label{mininet}
The first experiment we have carried out is to test and evaluate the performance of stealth TCP redirection. For this purpose, the Mininet virtualisation environment was used due to its ease of deployment and its good performance at network level. Once explained how the experiment has been carried out, details of the latency obtained by the attacker when performing the process explained in Section~\ref{redirection} will be given.

\subsubsection{Settings}

\begin{figure}[!h]
\centering
\includegraphics[width=9cm]{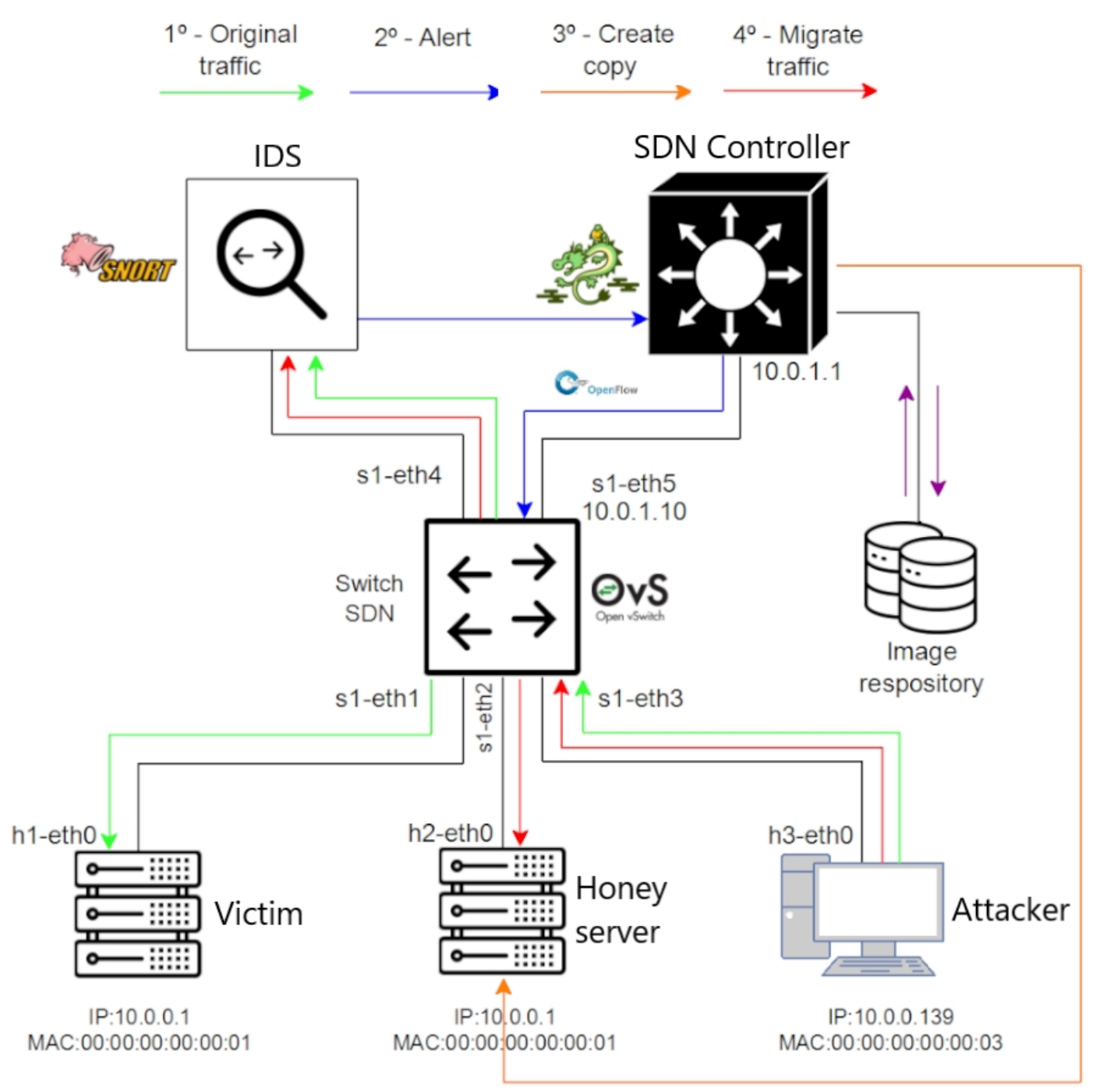}
\caption{Network topology for the proposed experiments.}
\label{a1}
\end{figure}

For this task, we set up the network topology shown in \figurename~\ref{a1}. For this Mininet experiment, we will perform steps 1, 2 and 4 as specified in the figure. We exclude the evaluation of honey server creation in this experiment; we focus only on testing stealth TCP redirection. Therefore, the honey server creation (step 3) is created statically before the process starts. 

Before running the topology, we defined the detection rule that will be used in Snort to initiate the redirection process reported in Listing~\ref{123}. In the rule, we specify that the migration occurs on packet number 100 from the client to the server. This rule is just an example to demonstrate our proposal and to generate an alert from the IDS.

\label{123}
\begin{lstlisting}[language=bash,caption={Snort rule}]
alert tcp any -&gt; X.X.X.X. 
any (msg: "MIGRATE"; flags: P.A.;
threshold: type threshold, track
by_dst, count 5, seconds 120; sid1000001;)
\end{lstlisting}

We can see in \figurename~\ref{a1} how the newly launched server has the same characteristics (i.e., IP and MAC address) so that the attacker does not recognise it, i.e., it is being deceived with a copied asset. Apart from this, the controller could seamlessly return the connection to the original server if Snort re-issued an alert to the controller. This ensured that the legitimate client connection would not be affected if the suspicious connection turned out to be a false positive, providing robustness to both the controller and this redirection technique.

\subsubsection{Results}

\begin{figure}[h]
\centering
\includegraphics[width=9cm]{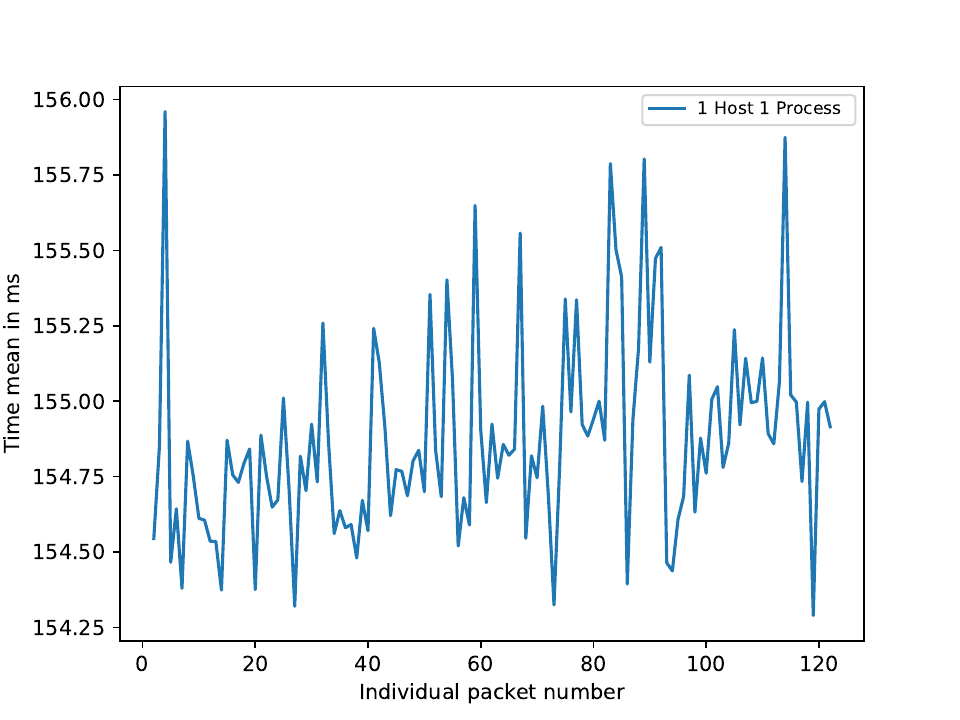}
\caption{Redirection latency means in the different experiments (1 Host - 1 Process).}
\label{r1}
\end{figure}

In \figurename~\ref{r1}, the average latencies experienced by the attacker when only redirection in a Mininet environment can be observed when the SDN controller handles only one connection in the experiments performed (100 times). In this figure, we can see how the latency remains constant during the whole experiment, without any change when performing the redirection in packet 100. These results suggest that the attacker does not perceive any suspicious actions caused by latency times, which makes the controller and the migration discrete. It is important to remark that, as seen in \figurename~\ref{r1}, the experiment terminates at packet number 120 because the redirection is done at packet 100, the subsequent latency times are not relevant for the evaluation of the redirection.

\noindent

Next, we analyse the same latency times when performing the migration in an environment where the controller is more saturated, i.e., with 20 simultaneous clients sending ICMP Request packets to other clients, i.e., ICMP Request packets are sent between these 20 clients. Each client will execute 70 infinite processes of sending ICMP Request packets, i.e., 70 processes will be executed with the \texttt{ping X.X.X.X.} instruction. In total, the SDN controller will have to moderate 1400 connections. With the 1400 packets, the controller will be saturated, it is then that we will carry out the attacker's execution, i.e., we will send TCP packets to the victim so that on the 100th packet we will be redirected.

\begin{figure}[h]
\centering
\includegraphics[width=9cm]{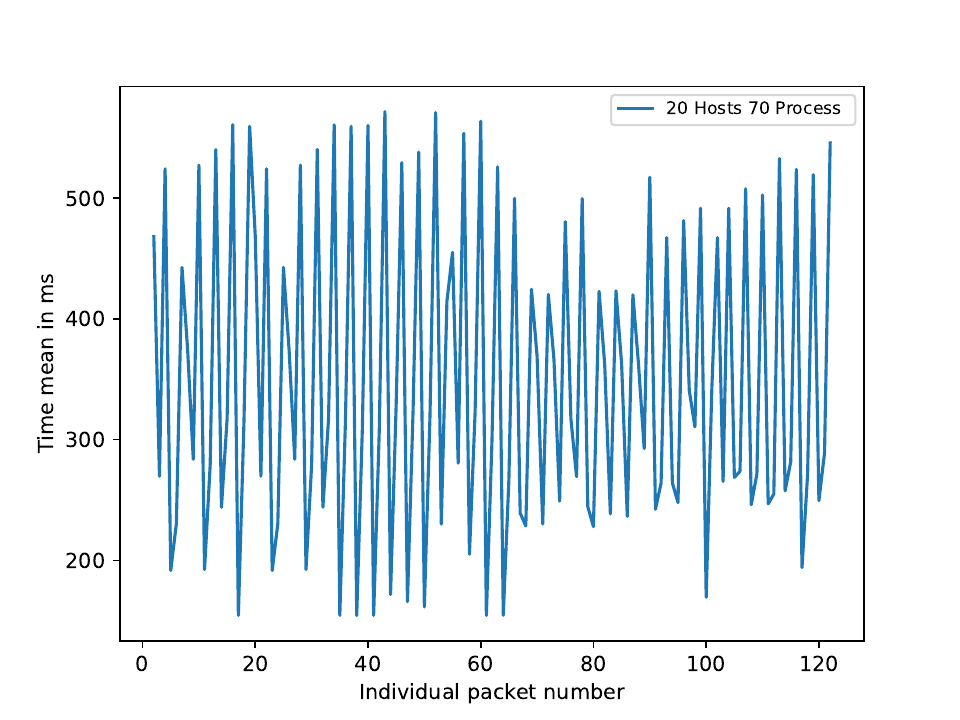}
\caption{Redirection latency mean in the different experiments (20 Host - 70 Process).}
\label{r2}
\end{figure}

In \figurename~\ref{r2}, the latencies experienced by the attacker in this scenario are presented. As in \figurename~\ref{r1}, we can observe that there is no significant difference between the packets before the number 100 and the packets after the number 100, which means that the attacker will not be able to detect when the redirection is taking place, as the connection remains stable.

With the results shown, we have demonstrated that the designed solution has the ability to efficiently deceive the attacker without being detected by the attacker. Furthermore, it has been proven that the saturation of the controller does not influence the stealth capacity of the proposed technique, thus increasing its defensive potential.

\noindent

\subsection{Redirection \& Copy on-demand}
\label{kvm}
Having tested and evaluated the stealth TCP redirection mechanism, we will now conduct an experiment that evaluates the full proposal, i.e., stealth TCP redirection to a honeypot created on demand. A key objective of this experiment is to significantly raise the level of realism of the system. This improvement is achieved by moving the Mininet environment to a more representative virtualised configuration. The need for this transition lies in replicating real-world conditions more accurately. By employing a virtualised environment, an additional degree of complexity and realism is introduced, allowing the performance of the system to be evaluated in a context closer to a practical deployment. This adjustment is essential to obtain more applicable and relevant results, as the fidelity of the test environment directly impacts the validity and usefulness of the conclusions drawn.

\subsubsection{Settings}

For this experiment, Kernel-based Virtual Machine (KVM) virtualisation technology has been used. This decision was made because of its ease of use from an Ryu SDN controller written in Python. 

The same architecture used in the Mininet experiment will be used, i.e., the one shown in \figurename~\ref{a1}. In this case, all 4 steps are executed since this experiment is going to carry out the copy on demand. Furthermore, the honey server will not be running as it will be created in real-time when the controller receives the alert from the IDS.

The same Snort rule will be used for this experiment to start the copy-on-demand and redirection process. On the other hand, the different virtual machines identified in the topology have been implemented and executed to carry out this experiment.

\subsubsection{Results}

We now proceed to analyse the latencies when, before performing the redirection, the controller creates a copy of the real-time asset.

\begin{figure}[h]
\centering
\includegraphics[width=9cm]{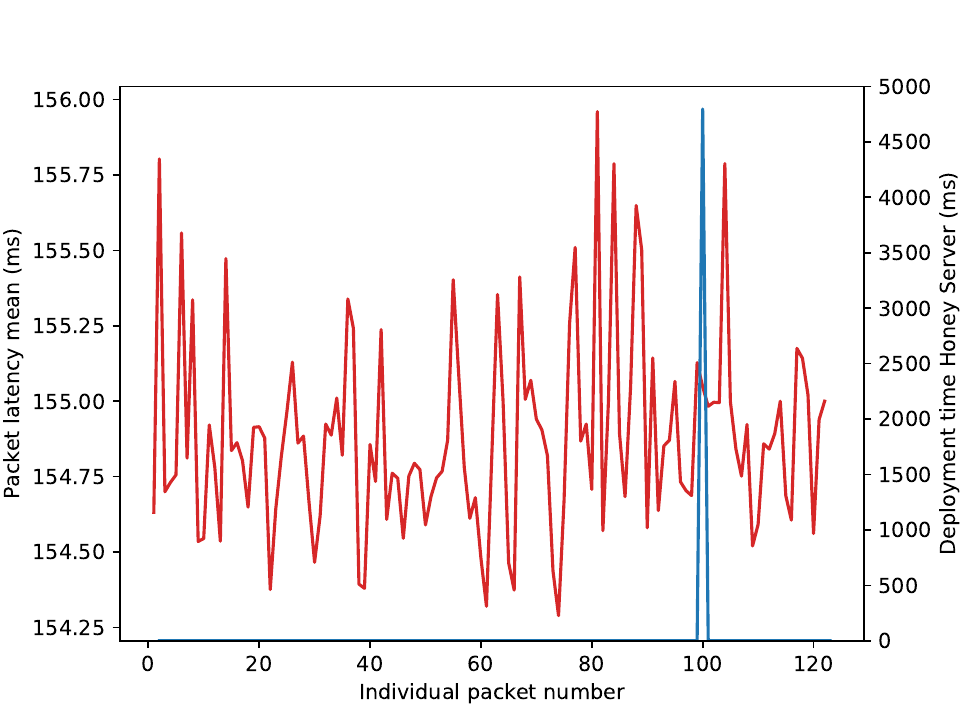}
\caption{Redirection \& copy on-demand latencies.}
\label{r3}
\end{figure}

\figurename~\ref{r3} presents the results of the experiment performed. In this graph, two different latencies are analysed: i) on the left, the latency obtained by the attacker in the exchange of messages with the victim/honey server, and ii) on the right, the latency time created when instantiating the honey server.

We observe how the redirection is carried out on the 100th packet of the connection, highlighting that there is no significant increase or decrease in latency, which deceives the attacker in a stealthy way. In addition, we notice a spike in latency when creating the honey server; however, this increase in latency is not noticeable to the attacker, as he/she is not redirected until the honey server instantiation is complete.

\noindent

\section{Conclusions and future work}
\label{conclusions}
This paper proposed, designed, implemented, and evaluated a CYDEC-based reaction method using stealthy TCP redirection to a honey asset created on demand. The implementation of the SDN controller and the TCP packet exchange during the connection were detailed, and the decisions made in creating the on-demand asset were explained.

The results obtained confirmed the system's performance and the effectiveness of the method in terms of its stealth capability, preventing the attacker from suspecting the connection. Therefore, an effective and dynamic deception method that could be used in various contexts for threat intelligence gathering and reaction was designed. This article will open the door to several new methods of defending against cyber threats by changing the defender's perspective. The defender, instead of blocking the attacker, the attacker will allow in and learned from him.

The next steps for the CYDEC theme, especially the part related to this work, involve integrating a threat information collection system and establishing an autonomous continuous learning system. Additionally, exploring testing scenarios in more realistic environments, such as mobile networks like 5G/6G, is essential.

\section*{ACKNOWLEDGMENTS}

This work has been partially funded by the strategic project DEFENDER from the Spanish National Institute of Cybersecurity (INCIBE) and by the Recovery, Transformation and Resilience Plan, Next Generation EU.

\section*{Biography}

\begin{figure}[!h]
    \begin{minipage}[H]{0.3\textwidth}
    \centering
        \includegraphics[width=0.7\textwidth]{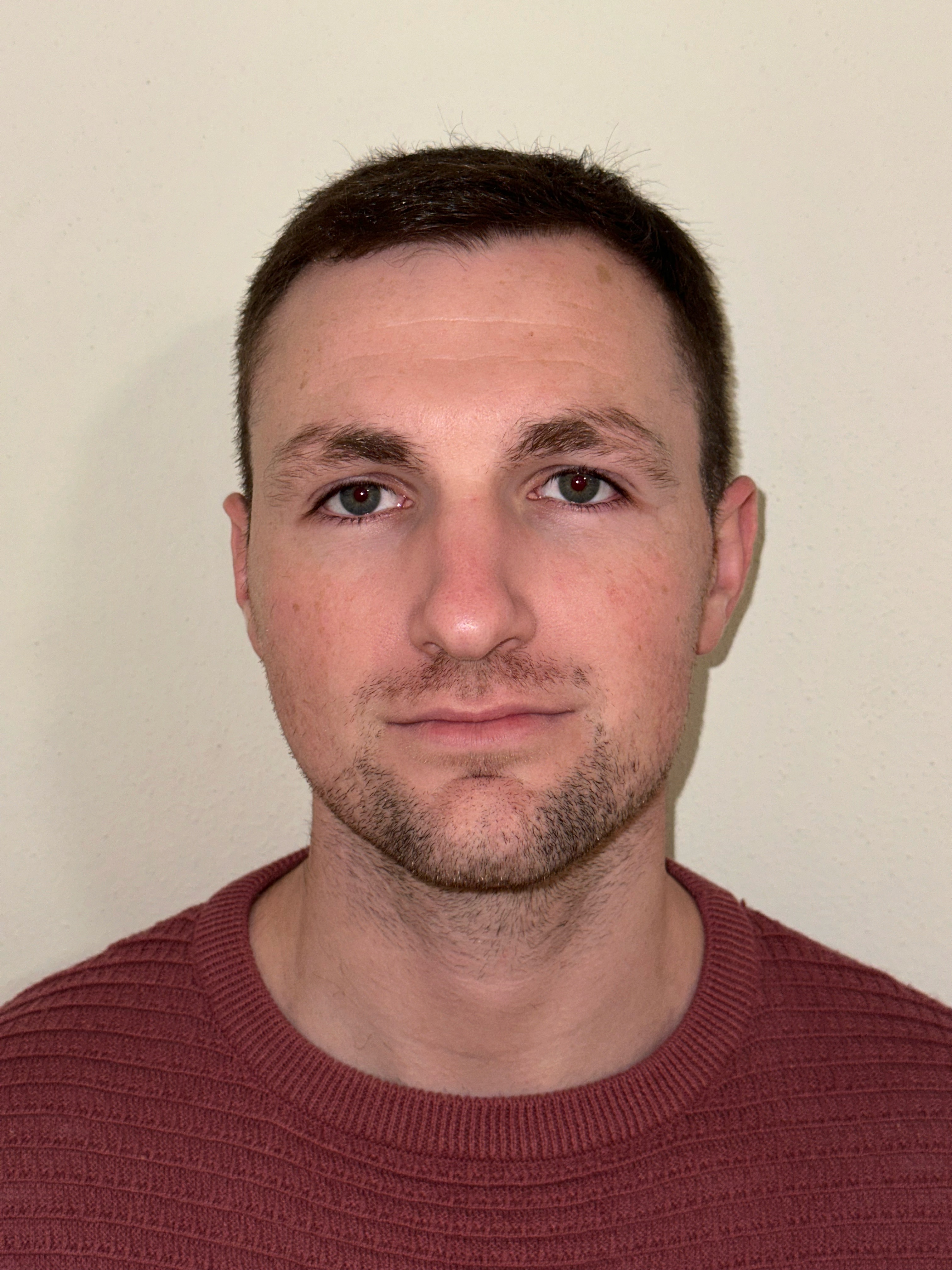}
    \end{minipage}
    \hfill
    \begin{minipage}[H]{0.7\textwidth}
        \textbf{Pedro Beltrán López} is Researcher of the CyberDataLab research group in the Department of Information and Communication Engineering of the University of Murcia, Murcia, Spain. His scientific activity is mainly focused on cybersecurity and artificial intelligence, with special emphasis on the use of artificial intelligence applied to on- techniques. Beltrán López received M.S.c degrees in Computer Science from the University of Murcia and is currently pursuing his Ph.D degrees.
    \end{minipage}
\end{figure}

\begin{figure}[!h]
    \begin{minipage}[H]{0.3\textwidth}
    \centering
        \includegraphics[width=0.8\textwidth]{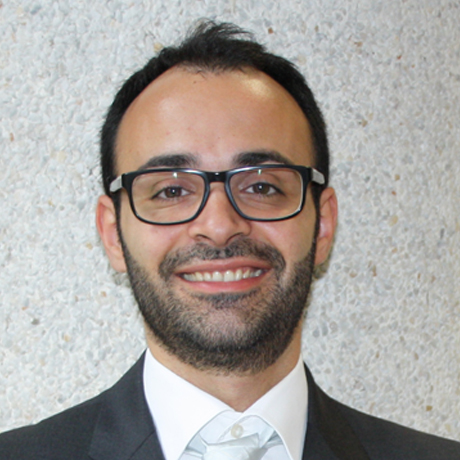}
    \end{minipage}
    \hfill
    \begin{minipage}[H]{0.7\textwidth}
        \textbf{Pantaleone Nespoli} is a postdoctoral researcher working together with the Department of Information and Communication Engineering at the University of Murcia, Spain, and the SCN team of the SAMOVAR laboratory, at Institut Polytechnique de Paris. His research is focused on cybersecurity and cyber defense training, with a particular interest in the detection and response to intrusions, and disinformation in social networks. Contact him at pantaleone.nespoli@um.es.
    \end{minipage}
\end{figure}

\begin{figure}[!h]
    \begin{minipage}[H]{0.3\textwidth}
    \centering
        \includegraphics[width=0.7\textwidth]{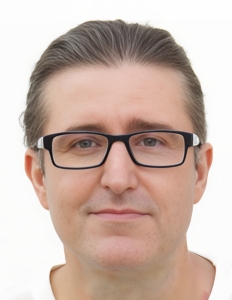}
    \end{minipage}
    \hfill
    \begin{minipage}[H]{0.7\textwidth}
        \textbf{Manuel Gil Pérez} is Associate Professor in the Department of Information and Communication Engineering of the University of Murcia, Murcia, Spain. His scientific activity is mainly devoted to cybersecurity, including intrusion detection systems, trust management, privacy-preserving data sharing, and security operations in highly dynamic scenarios. Gil Pérez received M.Sc. and Ph.D. degrees (later with distinction) in Computer Science from the University of Murcia.
    \end{minipage}
\end{figure}

\vfill

\end{document}